\documentclass{article}
\usepackage{arxiv}
\usepackage[utf8]{inputenc} 
\usepackage[T1]{fontenc}    
\usepackage{hyperref}       
\usepackage{url}            
\usepackage{booktabs}       
\usepackage{amsfonts}       
\usepackage{nicefrac}       
\usepackage{microtype}      
\usepackage{lipsum}		
\usepackage{graphicx}
\usepackage{natbib}
\usepackage{doi}
\usepackage{amsmath}
\usepackage{tabularx} 
\usepackage{url}

\title{Balancing Content Size in RAG-Text2SQL System}

\date{\today} 					

\author{ 
    {\hspace{1mm}Prakhar Gurawa} \\ 
    HP - Senior AI Engineer \\ 
    \texttt{prakhar.gurawa@hp.com} \\ 
    \And 
    {\hspace{1mm}Anjali Dharmik} \\ 
    HP - Senior AI Engineer \\ 
    \texttt{anjali.dharmik@hp.com} \\ 
}

\date{}

\renewcommand{\shorttitle}{Balancing Content Size in RAG-Text2SQL System}

\hypersetup{
pdftitle={Balancing Content Size in RAG-Text2SQL System},
pdfsubject={q-bio.NC, q-bio.QM},
pdfauthor={Prakhar Gurawa, Anjali Dharmik},
pdfkeywords={Text-to-sql , Retrieval Augmented Generation , Prompt Engineering , Large Language Models , Hallucination},
}

\begin{document}
\maketitle

\begin{abstract}
Large Language Models (LLMs) have emerged as a promising solution for converting natural language queries into SQL commands, enabling seamless database interaction. However, these Text-to-SQL (Text2SQL) systems face inherent limitations, hallucinations, outdated knowledge, and untraceable reasoning. To address these challenges, the integration of retrieval-augmented generation (RAG) with Text2SQL models has gained traction. RAG serves as a retrieval mechanism, providing essential contextual information, such as table schemas and metadata, to enhance the query generation process. Despite their potential, RAG + Text2SQL systems are susceptible to the quality and size of retrieved documents. While richer document content can improve schema relevance and retrieval accuracy, it also introduces noise, increasing the risk of hallucinations and reducing query fidelity as the prompt size of the Text2SQL model increases. This research investigates the nuanced trade-off between document size and quality, aiming to strike a balance that optimizes system performance. Key thresholds are identified where performance degradation occurs, along with actionable strategies to mitigate these challenges. Additionally, we explore the phenomenon of hallucinations in Text2SQL models, emphasizing the critical role of curated document presentation in minimizing errors. Our findings provide a roadmap for enhancing the robustness of RAG + Text2SQL systems, offering practical insights for real-world applications.
\end{abstract}

\keywords{Text-to-sql \and Retrieval Augmented Generation \and Prompt Engineering \and Large Language Models \and Hallucination}

\section{Introduction}
The rapid growth of data in relational databases has made accessing and analyzing this information increasingly challenging, especially for non-technical users who lack expertise in writing SQL queries. Translating natural-language queries into SQL commands is critical to enabling seamless interaction with complex databases. Text-to-SQL models address this need by leveraging natural language processing techniques to convert natural language into executable SQL queries, thus democratizing database access (\cite{shi:24}).

Large Language Models (LLMs) underpin many Text2SQL systems, offering impressive capabilities in understanding and generating natural language. However, they have limitations, such as hallucinations, outdated knowledge, and non-transparent reasoning processes (\cite{gao:24}). To overcome these challenges, Retrieval-Augmented Generation (RAG) has emerged as a transformative approach (\cite{lewis:20};\cite{izacard:22};\cite{guu:20};\cite{borgeaud:22}). By incorporating external knowledge from databases, RAG enhances the accuracy and reliability of LLMs for knowledge-intensive tasks. It enables continuous updates of knowledge and the integration of domain-specific information, integrating the intrinsic capabilities of LLMs with dynamic external data repositories (\cite{gao:24};\cite{wang:24}). The information retrieved from the RAG system serves as input for the LLM or Text2SQL model, forming a critical prompt component. The information retrieved from the RAG system is incorporated into the prompt provided to the LLM or Text2SQL model, playing a pivotal role in its ability to generate accurate SQL queries. The quality and quantity of this retrieved information not only determine the size of the prompt but also significantly impact the overall performance of the model. Thus, optimizing what and how much information is captured from the RAG system is crucial to achieving a balanced and efficient query generation process. In the broader domain of artificial intelligence, prompt engineering has emerged as a transformative discipline, harnessing the full potential of LLMs by tailoring input prompts to maximize their effectiveness and reliability (\cite{sahoo:24};\cite{chang:24}). 

Despite its advantages, RAG introduces its own set of challenges. The "Garbage In, Garbage Out" (GIGO) principle is highly applicable to the Retrieval-Augmented Generation (RAG) system, just as it is in other machine learning and data processing contexts. Incorporating excessive or irrelevant information can degrade the performance of LLM models, introduce noise into the prompts, and increase the risk of hallucinations (\cite{liu:23};\cite{maynez:20}). When provided with incorrect or incomplete information, LLMs may produce inaccurate responses or fail to comprehend the input query (\cite{bian:23};\cite{adlakha:24}). This makes the quality of retrieved content a critical factor in the overall performance of RAG-enabled systems.
In recent years, there has been a surge in the application of LLMs for Text2SQL tasks, driven by their enhanced performance, adaptability, and potential for future improvements (\cite{shi:24}). Although RAG + Text2SQL systems themselves are not enough to solve this problem, and in the real world, we might need other powerful additions in systems like TAG (\cite{biswal:24}) and many other additions, but we will keep our scope limited to RAG + Text2SQL. Real-world deployment of RAG + Text2SQL systems often demands careful optimization of document content used for retrieval. Larger documents may improve retrieval precision but simultaneously increase the risk of hallucinated SQL queries as prompts become longer and noisier. Balancing these factors is crucial to achieving reliable system performance.
This paper focuses on the following core challenges:
\begin{enumerate}
    \item The performance of RAG + Text2SQL systems is highly sensitive to the quality and size of the documents used for retrieval.
    \item Larger and detailed documents can improve the accuracy of RAG but exacerbate hallucination issues in Text2SQL models.
\end{enumerate}

Improving the RAG + Text2SQL system holds significant potential for enhancing efficiency in everyday tasks within the software industry.

\subsection{Why This Problem Matters}
In practical applications, such as business intelligence, automated reporting, and natural language interfaces for enterprise data, even a hallucinated SQL query can lead to incorrect insights and significant decision-making errors. Therefore, ensuring a balance between document size and quality is essential to maintaining the reliability of query results.
Enhancing the RAG + Text2SQL system holds significant potential for increasing efficiency in everyday tasks within the software industry. This, in turn, will contribute to the development of better software solutions, ultimately improving various aspects of our day-to-day lives.

\subsection{Objectives of the Paper}
This research addresses the trade-off between document size and quality in RAG + Text2SQL systems. Specifically, we:
\begin{enumerate}
    \item Analyse how varying document content impacts retrieval accuracy, hallucination rates, and system reliability.
    \item Explore the performance differences in RAG systems and RAG + Text2SQL systems across a range of document configurations.
    \item Propose a framework for designing concise, high-quality documents to achieve an optimal balance, minimizing hallucinations while maximizing retrieval effectiveness.
\end{enumerate}
By conducting extensive experiments with multiple document sets featuring varying levels and amounts of information, we provide actionable insights into optimizing RAG + Text2SQL frameworks for real-world applications.

\section{Related Work}
The field of Large Language Models (LLMs) is advancing at an unprecedented pace, with numerous innovations in architectures and methodologies to enhance their performance. One such approach is Retrieval-Augmented Generation (RAG), which has gained significant attention for improving the utility of LLMs by integrating domain-specific external retrieval mechanisms.

\subsection{Text-to-SQL Models}
Text-to-SQL models have evolved considerably over time, focusing on bridging the gap between natural language understanding and database querying. The survey by (\cite{shi:24}) explores the use of LLMs for Text-to-SQL tasks, highlighting the increasing importance of efficient querying in the face of growing data volumes. As relational databases rely on SQL for interaction, the need to democratize access for non-expert users has driven advancements in Text-to-SQL technologies. One notable work, Seq2SQL (\cite{zhong:17}), introduced a deep neural network architecture to translate natural language questions into SQL queries, addressing the accessibility challenge for users without SQL expertise.

\subsection{Impact of Document Characteristics on RAG Systems}
Recent studies have examined how document characteristics, including structure and content length, influence RAG system performance. The work by (\cite{soman:24}) provides insights into factors such as chunk size, embedding reliability, sentence- versus paragraph-based retrieval, keyword placement, and context order. These findings emphasize the importance of tailoring retrieval strategies to document properties, particularly for technical domains. Similarly, (\cite{zhao:24}) investigates the impact of retrieved document characteristics and prompt strategies on RAG system performance, highlighting how document quality, size, and content type significantly affect the accuracy and reliability of responses. This study further explores document selection methods and prompting strategies, underscoring their role in RAG + LLM frameworks.

\subsection{Prompt Design and Its Influence on LLM Performance}
Prompt engineering has emerged as a key factor in determining the performance of LLMs. Research such as (\cite{he:24}) reveals that prompt formatting significantly impacts GPT-based models (\cite{brown:20}), with no single format proving universally superior. This underscores the need for diverse prompt formats to enhance future LLM testing and performance optimization. Additionally, (\cite{kojima:23}) introduces Zero-shot-CoT, a zero-shot prompt designed to elicit chain-of-thought reasoning in LLMs, contrasting with previous few-shot approaches that require handcrafted examples. The work by (\cite{yugeswardeenoo:24}) explores question analysis prompting, demonstrating its potential to improve accuracy across diverse reasoning tasks, including mathematics and commonsense queries. These studies underline the critical role of prompt strategies in achieving robust and accurate model outputs, particularly in scenarios where document content directly influences prompt design, as in RAG + Text2SQL systems.

\subsection{Hallucination in LLMs and RAG Systems}
Hallucination remains a persistent challenge in generative AI and LLMs. Multiple studies (\cite{peng:23};\cite{dziri:21};\cite{feldman:23};\cite{Varshney:23};\cite{dhuliawala:23}) have investigated hallucination phenomena, analysed their origins and proposed mitigation strategies. A notable contribution comes from (\cite{jesson:24}), which presents a technique for estimating hallucination rates in conditional generative models within in-context learning frameworks. These findings are crucial for our research, as document quality and retrieved information significantly impact hallucination rates in RAG + Text2SQL systems.

Building upon the existing body of work, our research focuses on the interplay between document size and quality in RAG + Text2SQL systems. By analysing how these factors affect retrieval accuracy, prompt design, and hallucination rates, we aim to provide actionable insights for optimizing these systems in real-world applications.

\section{Problem Definition}
When integrated with text-to-SQL (Text2SQL) models, the effectiveness of retrieval-augmented generation (RAG) systems heavily depends on the quality and size of the documents used for retrieval. This dependency arises from these documents' role in providing schema and context for accurate SQL query generation.

Our study used a subset of the SPIDER data set (\cite{yu:2019}), a widely recognized benchmark for Text2SQL systems, as the basis for experimentation. By sampling various SQL table schemas from the SPIDER data set, we constructed multiple document representations to systematically evaluate their influence on the combined RAG + Text2SQL system.

Each document in our study corresponds to a single table and includes: 
\begin{enumerate}
    \item The table's schema includes column names and other relationships among columns.
    \item Additional metadata that may aid the RAG system's understanding and retrieval accuracy.
\end{enumerate}

To explore the impact of document content, we created multiple iterations of data sets based on the SPIDER data set. These iterations varied in document size, level of detail, and quality, enabling us to analyze their effect on the RAG system and the RAG + Text2SQL system as a unified framework. The primary goal of this study is to determine the optimal document structure that balances two competing objectives:
\begin{enumerate}
    \item Minimizing hallucination risks in Text2SQL outputs caused by noisy or excessive document content.
    \item Maximizing retrieval relevance and accuracy in SQL query generation by ensuring sufficient and precise document information.
\end{enumerate}
By conducting controlled experiments across these varied document representations, we aim to provide actionable insights into document design strategies for RAG + Text2SQL systems. This research highlights how document quality and content size directly influence system performance, paving the way for more robust and reliable implementations of such frameworks in real-world applications.

\section{Experimental Setup}
In this section, we present the key components of our experimental setup \footnote{All the experiment code is presented at the link : \url{https://github.com/prakhargurawa/Balancing-Content-Size-In-RAG-Text2SQL-System}.}. We begin by introducing the data set employed to evaluate the performance of the RAG+Text2SQL system. Next, we outline the document creation process tailored for the RAG system, followed by a detailed discussion of the RAG system architecture and its configuration parameters. Finally, we describe the Text2SQL model, an advanced LLM-based system, elaborating on its prompt design and implementation strategy.

\subsection{About Data Set: Spider}
The SPIDER data set \footnote{The data set has been downloaded by following link : \url{https://yale-lily.github.io/spider}.}, introduced in "Spider: A Large-Scale Human-Labeled data set for Complex and Cross-Domain Semantic Parsing and Text-to-SQL Task" (\cite{yu:2019}) is a benchmark designed to evaluate the ability of models to generate SQL queries from natural language queries across diverse domains. It consists of 10,181 questions and 5,693 unique complex SQL queries, spanning 200 databases with multiple tables and representing 138 different domains. The data set's complexity and diversity make it a valuable resource for advancing research in semantic parsing and text-to-SQL systems. The data set is highly versatile and includes SQL queries across multiple levels of complexity, encompassing all major SQL query components. It features queries with SELECT clauses involving multiple columns and aggregations, along with WHERE, GROUP BY, HAVING, ORDER BY, and LIMIT clauses. Advanced operations such as JOIN, INTERSECT, EXCEPT, UNION, NOT IN, OR, AND, EXISTS, and LIKE are well-represented. Additionally, the data set includes nested queries and subquery structures, reflecting real-world query scenarios. To ensure comprehensive coverage, the annotators have carefully ensured that every table in the database is referenced in at least one query (\cite{yu:2019}).

The data set is structured into training, validation, and test subsets to facilitate systematic evaluation. We utilized a curated section of the data set for our experiments, focusing on 15 domains. This subset comprises 719 queries and includes 54 tables, enabling us to conduct domain-specific analysis while maintaining a manageable scale for experimentation. Table 1 presents an overview of the 15 distinct domains covered in the data set, along with the corresponding SQL queries. The data set comprises 719 SQL queries, highlighting its diverse and comprehensive coverage across various database schemas.

\begin{table}
	\caption{Distribution of SQL queries over different selected domains selected for benchmarking}
	\centering
	\begin{tabular}{lll}
		\toprule
		Domain     & No. of SQL queries      \\
		\midrule
		farm & 40 \\ 
film\_rank & 68 \\ 
election & 40 \\ 
wrestler & 20 \\ 
wedding & 30 \\ 
swimming & 30 \\ 
climbing & 40 \\ 
device & 40 \\ 
loan\_1 & 80 \\ 
movie\_1 & 98 \\ 
railway & 21 \\ 
coffee\_shop & 18 \\ 
game\_1 & 86 \\ 
insurance\_policies & 48 \\ 
product\_catalog & 42 \\ 
		\bottomrule
	\end{tabular}
	\label{tab:table}
\end{table}

\subsection{Document Variations}
RAG systems retrieve information by analysing the similarity between the user query and the available documents. Based on this similarity, the system identifies and selects the top k most relevant documents, accompanied by their respective similarity scores, ensuring that the retrieved content is closely aligned with the user's query. In our experiment, each document represents a single SQL table, encapsulating information about the table schema and, in some cases, additional metadata such as textual descriptions or example insert queries. The documents were created by sampling table schemas from the SPIDER data set, generating variations based on schema representation and adding contextual details, as outlined in the previous section.

When a user provides a natural language query, it is first passed through the RAG system, which retrieves the top three relevant documents from the collection. These documents and RAG are expected to represent the tables most pertinent to answering the query. The retrieved documents are combined with the original user query and specific Text2SQL instructions to form a prompt. This prompt is provided to the Text2SQL model, which processes the input and generates the desired SQL query. This setup ensures that the system leverages both the retrieval capabilities of RAG and the reasoning power of the Text2SQL model. By evaluating the performance across various document representations, we aim to understand how changes in document quality and content size influence retrieval accuracy, prompt construction, and the final SQL query generation.

To explore the effect of document quality and content on RAG + Text2SQL performance, we created several variations of table schema documents:
\begin{enumerate}
    \item Spider-Data-1: Includes only the original CREATE TABLE statements from the SPIDER data set.
    \item Spider-Data-2: Features a uniform representation of the CREATE TABLE statements across all tables to ensure consistency.
    \item Spider-Data-3: Contains the modified CREATE TABLE statements along with one example INSERT query to illustrate data usage.
    \item Spider-Data-4: Extends Spider-Data-3 by including two example INSERT queries to provide richer context.
    \item Spider-Data-5: Combines the modified CREATE TABLE statements with a textual description of the table and its columns.
    \item Spider-Data-6: Builds upon Spider-Data-5 by adding one example INSERT query to the document.
    \item Spider-Data-7: Extends Spider-Data-6 by including two example INSERT queries for even greater detail.
\end{enumerate}

\subsubsection{Document Quality vs. Size Trade-off}
The variations in document content among the different document set reflect a trade-off between document size and quality, a critical factor in the RAG + Text2SQL pipeline. It will impact the overall system in the following ways: 
\begin{enumerate}
    \item Effect on Retrieval: Larger documents, enriched with more contextual and descriptive information, enhance the retrieval step by providing more relevant matches to the user query. For instance, adding textual descriptions or sample data inserts provides the RAG system with richer signals to determine relevance.
    \item Effect on Prompt Size: The retrieved documents, combined with the user query and predefined instructions, are formatted into prompts for the Text2SQL model. Large documents can increase prompt size, which, while beneficial for retrieval, risks overloading the Text2SQL model. This can lead to inefficiencies in token usage and increased chances of hallucination, where the model generates queries based on incorrect or fabricated interpretations of the input.
    \item Balancing Trade-offs: This study hypothesizes that balancing document size and quality is essential for optimizing overall system performance. Documents should contain enough information to support accurate retrieval without overburdening the Text2SQL model's processing capabilities.
\end{enumerate}

\subsection{Description of RAG system}
Retrieval-Augmented Generation (RAG) introduces an information retrieval process that enhances the generative model's accuracy and robustness by fetching relevant objects from external data stores. This integration allows RAG systems to dynamically incorporate up-to-date and domain-specific knowledge, significantly improving their performance, particularly in knowledge-intensive tasks (\cite{zhaozhang:24}).

A retrieval mechanism is embedded into the model pipeline in an RAG system. It fetches contextually relevant information from an external knowledge base or document corpus based on the user query. The retrieved content is then combined with the original query and passed to a generative model, which uses this enriched context to produce its output. This approach represents a transformative shift in Generative AI, creating more transparent ("glass-box") models that excel in accuracy and reliability, especially in domains requiring precise information (\cite{khan:24}). RAG systems also mitigate the need for frequent retraining of large models, reducing both computational and financial costs. This adaptability makes RAG particularly appealing for enterprise applications, where maintaining up-to-date models is essential. 

We designed our RAG system using the following components:
\begin{enumerate}
    \item Framework: We utilized LangChain (\cite{langchain}), a robust framework that simplifies the development of advanced applications integrating language models. Its modular design allows seamless integration of RAG components.
    \item Embeddings: Semantic search in RAG systems relies on vector embeddings. For our implementation, we used all-MiniLM-L12-v2, a sentence-transformer model capable of converting textual data into fixed-size embeddings. This model is ideal for clustering and semantic search tasks and has demonstrated superior performance among open-source embedding models (\cite{Roman:24}).
    \item Vector Store: Efficient storage and querying of embeddings are essential for the RAG pipeline. We employed FAISS (Facebook AI Similarity Search) (\cite{douze:24}), an open-source library optimized for fast and lightweight similarity searches. FAISS retrieves relevant document chunks during query processing with high precision.
    \item k Value (Number of Retrieved Documents): The k value in RAG systems specifies the number of documents retrieved from the external knowledge source for a given query. These documents provide context to the generative model for response generation. In our setup, we set k = 3. This choice aligns with our task of SQL query generation, where a single SQL query typically involves no more than three tables, each represented as a document in our system. Selecting an appropriate k value is crucial as it directly impacts system performance. A higher k value offers a more comprehensive context, potentially improving response quality. However, it can also introduce noise or irrelevant information, increasing processing complexity and a higher risk of hallucinations or incorrect outputs. Conversely, a smaller k value may provide insufficient context, resulting in incomplete or inaccurate responses. Determining the optimal k value requires empirical evaluation to balance sufficient context with minimizing irrelevant information.
\end{enumerate}
This carefully constructed RAG system serves as the foundation for our experiments, enabling us to evaluate the impact of document size and quality on performance in RAG + Text2SQL settings.

\subsection{About Text2SQL Model}
Language models are transforming data management by enabling users to query databases using natural language, eliminating the need for specialized SQL knowledge. This innovation has spurred extensive research in both Text2SQL and Retrieval-Augmented Generation (RAG) methods (\cite{biswal:24}). For our Text2SQL tasks, we utilized the LLama-3-based SQLCoder-8B (\cite{llama-3-sqlcoder-8b:24}), a state-of-the-art AI model designed to translate natural language queries into SQL. SQLCoder-8B addresses the limitations of traditional Text2SQL models, setting new benchmarks for accuracy, adaptability, and ease of use. With an impressive accuracy rate exceeding 90\% in zero-shot scenarios, the model demonstrates superior performance across diverse query contexts, making it a cornerstone of our study.

To ensure optimal performance, we carefully tuned several hyperparameters during experimentation:
\begin{enumerate}
    \item Temperature: Set to 0.01 to minimize output randomness, enhancing generated queries' consistency.
    \item Top\_p: Configured to 0.7 to control the diversity of generated responses while maintaining relevance.
    \item Max New Tokens: Limited to 1024, ensuring the model generates complete and syntactically correct SQL queries without unnecessary verbosity.
    \item Return Full Text: Set to False to streamline outputs for integration into downstream processes.
\end{enumerate}
These optimizations were critical in enabling SQLCoder-8B to function effectively within the RAG + Text2SQL system, providing accurate and efficient query generation. The model's robustness and adaptability make it an ideal choice for exploring the impact of document size and quality on system performance.

Figure 1 illustrates the complete workflow of our Retrieval-Augmented Generation (RAG) integrated Text2SQL system. It showcases the interaction between the user query, the retrieval process, and the generative model, highlighting how relevant context from external documents is seamlessly incorporated into the SQL generation process.

\begin{figure}
    \centering
    \includegraphics[width=1\textwidth]{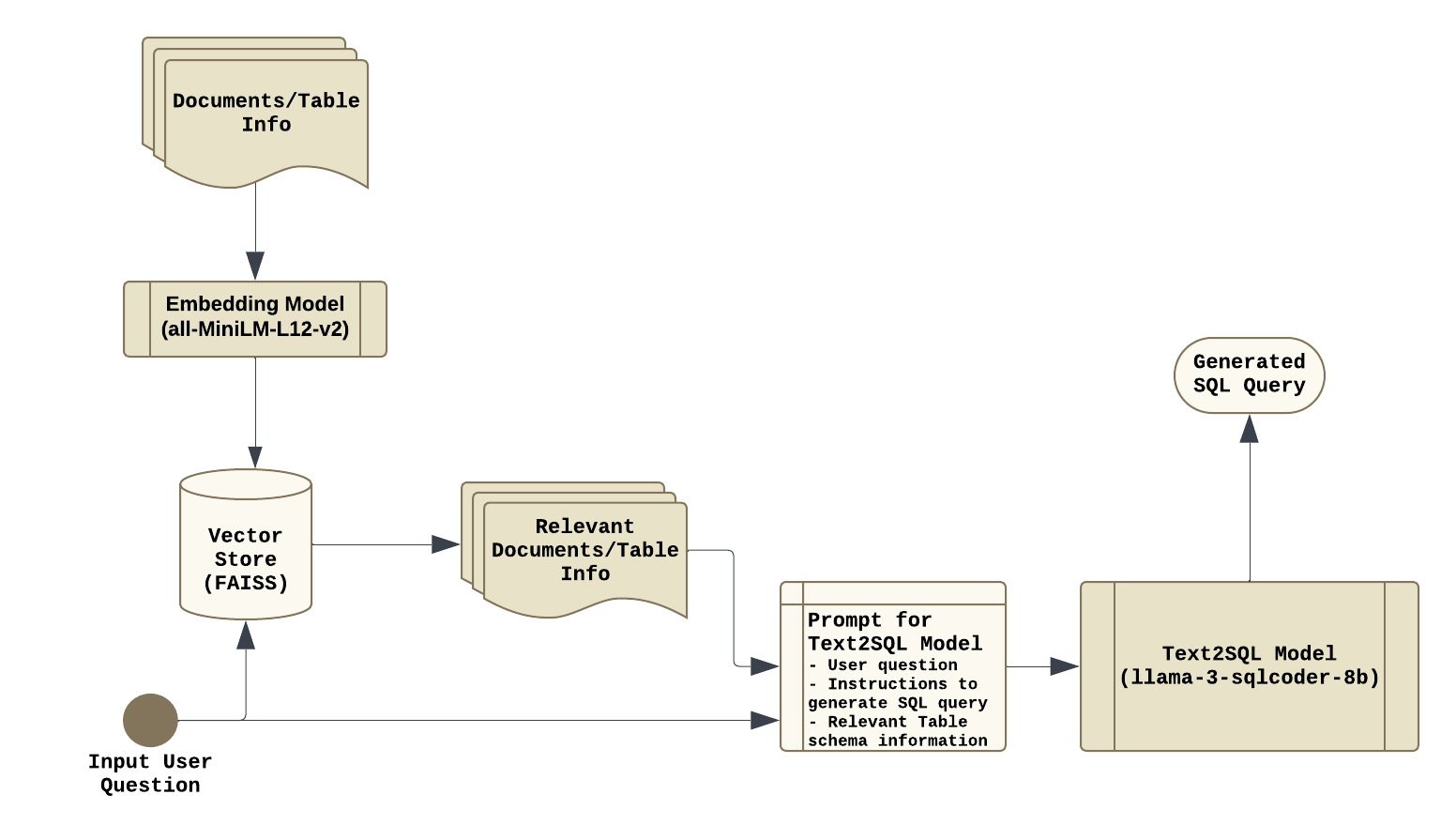}
    \caption{End-to-End Pipeline of the RAG + Text2SQL System}
    \label{fig:example_image}
\end{figure}

\subsubsection{Prompt Design for the Text2SQL Model}
One of the key factors influencing the performance of large language models (LLMs) is the structure and content of the input prompt. Each model requires a carefully crafted prompt format to ensure optimal functioning. In the case of our Text2SQL model, the prompt is designed to integrate essential components that guide the model in generating accurate SQL queries. The prompt consists of the following three core elements:

\begin{enumerate}
    \item User Question: This is the user's natural language query, which needs to be converted into a corresponding SQL statement.
    \item Instructions: These are specific directives tailored to instruct the Text2SQL model on how to process the input and generate the desired SQL query. Clear and precise instructions play a pivotal role in ensuring the model adheres to the intended logic and query format.
    \item Information from the RAG System: In our system, the Retrieval-Augmented Generation (RAG) mechanism provides critical context by retrieving schema details and relevant knowledge from the top 3 most relevant documents or tables. This retrieved information is incorporated into the prompt, enabling the Text2SQL model to understand the database structure and generate accurate SQL queries.
\end{enumerate}

Below, we present an example of the prompt structure utilized in our Text2SQL model. The example illustrates how instructions are provided to guide the model in generating an SQL query, along with the user's question and the corresponding table schema.

\begin{verbatim}
<|begin_of_text|><|start_header_id|>user<|end_header_id|>
Generate a SQL query to answer this question: `{question}`

### Instructions
- Given an input question, create a syntactically correct query to run, then 
look at the results of the query and return the answer.
- Never query for all the columns from a specific table; only ask for the 
relevant columns given the question.
- Only return the columns the user asks for; do not give any additional ID 
column that the user does not ask for explicitly.
- Do not add ORDER BY in the query if the user has not explicitly asked to 
order it.
- If you cannot answer the question with the available database schema, 
return 'I do not know.'
- Make sure that you never return two columns with the same name, especially 
after joining two tables. You can differentiate the same column name by 
applying column_name + table_name.
- DO NOT make any DML statements (INSERT, UPDATE, DELETE, DROP, etc.) to the 
database.
- If you are fetching data from a table only then use its columns to filter 
out the data.
- You MUST double-check your query before executing it. If you get an error 
while executing a query, rewrite the query and try again.

DDL statements:
CREATE TABLE "farm" (
"Farm_ID" int,
"Year" int,
"Total_Horses" real,
"Working_Horses" real,
"Total_Cattle" real,
"Oxen" real,
"Bulls" real,
"Cows" real,
"Pigs" real,
"Sheep_and_Goats" real,
PRIMARY KEY ("Farm_ID")
);

CREATE TABLE "farm_competition" (
"Competition_ID" int,
"Year" int,
"Theme" text,
"Host_city_ID" int,
"Hosts" text,
PRIMARY KEY ("Competition_ID"),
FOREIGN KEY (`Host_city_ID`) REFERENCES `city`(`City_ID`)
);


CREATE TABLE "competition_record" (
"Competition_ID" int,
"Farm_ID" int,
"Rank" int,
PRIMARY KEY ("Competition_ID","Farm_ID"),
FOREIGN KEY (`Competition_ID`) REFERENCES `farm_competition`(`Competition_ID`),
FOREIGN KEY (`Farm_ID`) REFERENCES `farm`(`Farm_ID`)
);<|eot_id|><|start_header_id|>assistant<|end_header_id|>
The following SQL query best answers the question `{question}`:
```sql
\end{verbatim}

\section{Evaluation Metrics}
To evaluate the performance of the RAG system and the combined RAG + Text2SQL system, we have created multiple sets of documents. Our goal is to assess how document content and quality variations impact both systems' performance. By analyzing these variations, we aim to understand how different document characteristics influence the overall performance of the RAG-based retrieval process and the subsequent Text2SQL query generation.

\subsection{Evaluation Metric for the RAG System}
The RAG system evaluates a user query by retrieving the top-k relevant documents and their corresponding relevance scores, where a lower score indicates higher relevance. To assess the RAG system's performance on the different data sets created from the SPIDER data set, we measure and compare its ability to effectively distinguish relevant tables from non-relevant ones.

A superior RAG system is characterized by its ability to assign a broader range of relevance scores, demonstrating a clear differentiation between valid and non-relevant documents. Conversely, if the scores for the top-k documents are closely clustered, it indicates difficulty in discrimination and potential confusion in the system. We executed all user queries through the RAG system for each data set variation and recorded the top-k retrieved documents along with their associated relevance scores.
The distribution of these scores was analyzed using the following metrics:

\begin{enumerate}
\item Discounted Cumulative Gain (DCG): Evaluates the relevance of retrieved documents, emphasizing higher-ranked documents.

       \begin{equation*} 
            DCG_p = \sum_{i=1}^{p} \frac{rel(i)}{\log_2(i+1)}
        \end{equation*}

    \text{where:} 
    \begin{itemize}
        \item $rel(i)$ is the relevance score of the document at position $i$.
        \item $p$ is the rank position (top-$k$ in your case).
    \end{itemize}

\item Standard Deviation (Std Dev): Measures the spread of relevance scores to understand the variability in the system's differentiation capabilities.

    \begin{equation*} 
        StdDev = \sqrt{\frac{1}{k} \sum_{i=1}^{k} (rel(i) - \overline{rel})^2}
    \end{equation*}

    \text{where:}
    \begin{itemize}
        \item $k$ is the number of top documents retrieved.
        \item $rel(i)$ is the relevance score of the document at position $i$.
        \item $\overline{rel}$ is the mean relevance score of the top-$k$ documents:
            \begin{equation*}
                \overline{rel} = \frac{1}{k} \sum_{i=1}^{k} rel(i)
            \end{equation*}
    \end{itemize}

\item Range: Captures the difference between the highest and lowest scores among the top-k documents to indicate the breadth of the system's scoring distribution.
    \begin{equation*}
        Range = \max(rel(i)) - \min(rel(i))
    \end{equation*}

    \text{where:}
    \begin{itemize}
        \item $\max(rel(i))$ is the highest relevance score among the top-$k$ documents.
        \item $\min(rel(i))$ is the lowest relevance score among the top-$k$ documents.
    \end{itemize}
\end{enumerate}

For each data set, we computed the average values of these metrics across all user queries. By comparing the aggregated metrics, we aim to evaluate and rank the performance of different RAG system configurations and document representations. This approach ensures a comprehensive understanding of the RAG system's ability to retrieve and rank relevant tables effectively across varying data set structures.

\begin{align*}
  \text{AvgDCG} &= \frac{1}{Q} \sum_{q=1}^{Q} \text{DCG}_q \\ 
  \text{AvgStdDev} &= \frac{1}{Q} \sum_{q=1}^{Q} \text{StdDev}_q \\
  \text{AvgRange} &= \frac{1}{Q} \sum_{q=1}^{Q} \text{Range}_q
\end{align*}

\text{Where } Q \text{ is the total number of user queries.}

\subsection{Evaluation Metrics for the RAG + Text2SQL System}
In this research, our primary focus is not on measuring the performance of the Text2SQL models in isolation, as that would require experiments with different hyperparameters and model configurations, which are beyond the scope of this study. Instead, our objective is to evaluate the performance of various document data sets on the RAG + Text2SQL system. To achieve this, we assess the quality of the SQL queries generated by the whole RAG + Text2SQL system.

The evaluation involves measuring instances of hallucination and conducting SQL query similarity checks. For this purpose, we compare the generated SQL queries with the corresponding correct queries provided in the SPIDER data set. The following metrics were utilized to measure query similarity and identify discrepancies:

\begin{enumerate}
    \item Normalized Edit Distance: The process involves following steps, calculate the edit distance between the generated and correct SQL queries. Then, normalize the distance by dividing it by the length of the queries. If the normalized edit distance is below a threshold of 0.5 (determined empirically), the queries are considered similar; otherwise, they are not.
     \item Embedding Matching: The steps are converting SQL queries into textual embeddings using the all-MiniLM-L12-v2 model. Then, the cosine similarity between the embeddings is measured. Finally, queries with a similarity score less than or equal to 0.85 are classified as mismatched.
    \item Fuzzy Matching: Involves using fuzzy string-matching algorithms to calculate the similarity score between the queries. A threshold of 75 was set, so queries with a similarity score equal to or exceeding this value are considered similar.
    \item SQL Component Matching: It involves performing a component-wise comparison between the generated and correct SQL queries, evaluating: 
    \begin{itemize}
    \item Table Selection: Matches the tables involved in the query.
    \item Column Selection: Matches the columns being queried.
    \item Operation Selection: Matches the operations (e.g., JOIN, WHERE, GROUP BY) used in the query.
    \end{itemize}
    Any mismatches in these components indicate that the generated query does not align with the expected query, highlighting instances of hallucination in query generation.
    \item Database Execution Comparison: In this step, execute the generated and correct SQL queries on a PostgreSQL database and compare the outputs. Then, it utilizes exact matching and rule-based matching of results. Mismatches in the outputs suggest that the generated SQL query is either incorrect or unsuitable for the given query. 
\end{enumerate}
Finally, we measure the percentage of SQL queries showing hallucination as the above metrics. By applying these metrics, we systematically evaluate the impact of document variations on the RAG + Text2SQL system and provide insights into the quality and reliability of the generated SQL queries.

\section{Results and Analysis}
To evaluate the impact of varying document content and quality on RAG + Text2SQL systems, we utilized seven document sets derived from the SPIDER data set. The data sets ranged from Spider-Data-1, containing only the original CREATE TABLE statements, to Spider-Data-7, which incorporated modified CREATE TABLE statements, textual table descriptions, and two example INSERT queries to provide comprehensive context. The progression from Spider-Data-1 to Spider-Data-7 systematically increased the amount of data and contextual richness available to the RAG system. The primary objective was to analyze how the performance of both the RAG system and the combined RAG + Text2SQL system changed as the content in the documents increased. While all additional information across data sets, such as textual descriptions and example INSERT queries, was relevant, we aimed to determine the effect of such content augmentation on retrieval relevance and SQL query generation. The added information aimed to improve the Text2SQL model's understanding of tables, schemas, and data types.

\subsection{RAG Retrieval Performance}
To assess the RAG system's performance, we analyzed the relevance scores assigned to the top-k documents retrieved for each user query. A superior RAG system is characterized by a broader spread of relevance scores, indicating its ability to effectively distinguish between relevant and irrelevant documents.

Key metrics were used to capture the performance:
\begin{enumerate}
    \item Range of Relevance Scores: The difference between the highest and lowest relevance scores for the top-k documents.
    \item Standard Deviation of Relevance Scores: Indicates the variability in scores and the system's differentiation capability.
    \item Discounted Cumulative Gain (DCG): Measures the ranking quality of retrieved documents, with higher emphasis on documents ranked at the top.
\end{enumerate}

The averages of these metrics were calculated across all queries and plotted in Figures 2, 3, and 4. These figures illustrate how the performance metrics varied with the increasing document content from Spider-Data-1 to Spider-Data-7.

\begin{figure}
    \centering
    \begin{minipage}[t]{0.48\textwidth}
        \centering
        \includegraphics[width=\textwidth]{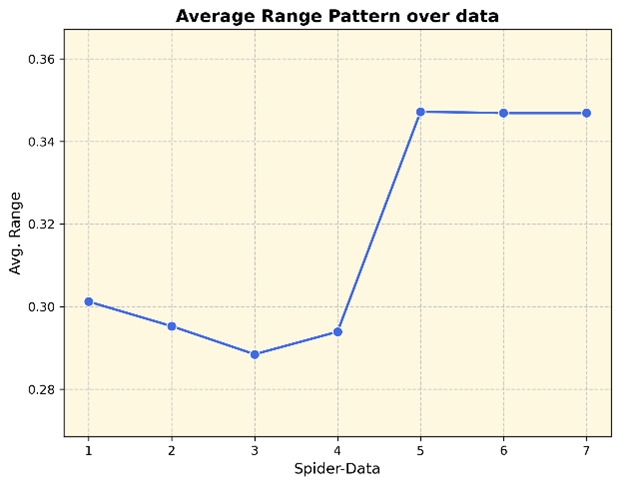}
        \caption{Variation of Avg. Range of Scores over different document sets}
        \label{fig:avg_range_pattern}
    \end{minipage}%
    \hfill
    \begin{minipage}[t]{0.48\textwidth}
        \centering
        \includegraphics[width=\textwidth]{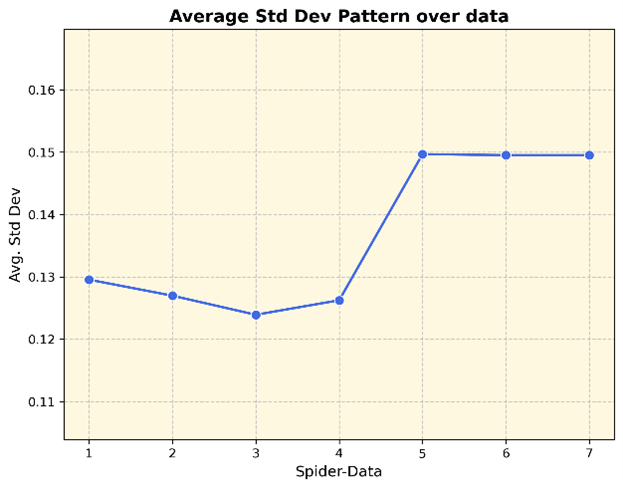}
        \caption{Variation of Avg. Std Dev of Scores over different document sets}
        \label{fig:avg_std_dev_pattern}
    \end{minipage}
    
    \vspace{1em} 

    \begin{minipage}[t]{0.6\textwidth}
        \centering
        \includegraphics[width=\textwidth]{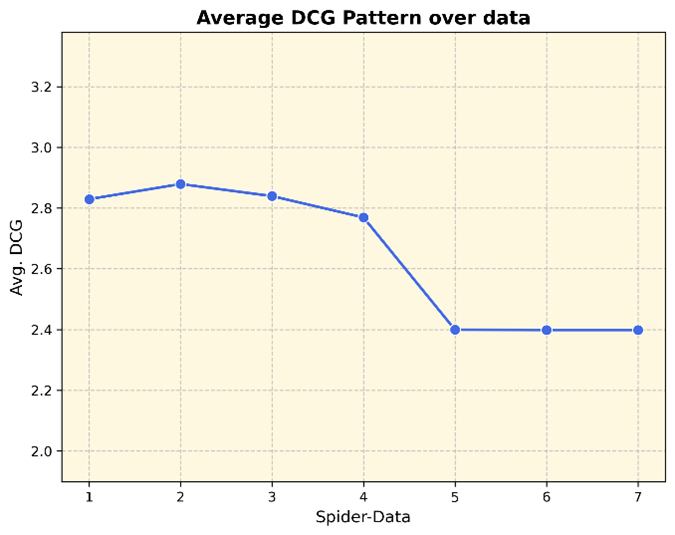}
        \caption{Variation of Avg. DCG of Scores over different document sets}
        \label{fig:avg_dcg_pattern}
    \end{minipage}
\end{figure}

The following observations have been derived from the above figures regarding the performance of the scores and the RAG system:
\begin{enumerate}
    \item Range of Scores: The range increased with richer document content, reflecting improved differentiation between relevant and non-relevant documents by the RAG system.
    \item Standard Deviation: A higher standard deviation was observed for data sets with richer content, supporting that detailed documents enhanced the system's discrimination capability.
    \item DCG: The DCG metric demonstrated a downward trend as more information was added, indicating that the top-ranked documents retrieved by the RAG system became increasingly relevant in the DCG formula; we used the RAG score, which the lower they are, the more relevant the documents are.
    \item The plots illustrate a significant improvement in the RAG system's performance, beginning with Document Data set 5, which includes textual descriptions of table schemas. This enhancement underscores the value of providing additional contextual information about SQL table schemas, enabling the RAG system to better differentiate between tables and rank them more effectively based on relevance. Document sets with richer content, such as Spider-Data-6 and Spider-Data-7, demonstrated the performance of RAG due to better contextual information about table schemas and data types.
    \item Also, adding the INSERT statement in the document files did help in document sets 3 and 4, where the content itself was less, but as the amount of information increased from document set 5, adding the INSERT statement in 6 and 7 didn't help the RAG much.
\end{enumerate}

These results highlight the relationship between document quality and retrieval effectiveness. Comprehensive documents with structured information enhance the RAG system's ability to retrieve highly relevant content, which directly impacts the Text2SQL system's performance.

\subsection{RAG + Text2SQL Performance}
The progression of document content across data sets in our experiments directly influenced the prompt size for the Text2SQL model. While richer document content provides more context, enabling the Text2SQL model to better understand table schemas and generate accurate SQL queries, the larger prompt size also increases the likelihood of hallucinations. Striking a balance between providing comprehensive information and maintaining prompt clarity is critical for optimizing system performance.

To evaluate the impact of document variations on the RAG + Text2SQL system:
\begin{enumerate}
    \item Query Generation: The system was run for every natural language query in all seven document sets. The Text2SQL model generated SQL queries for each query.
    
    \item Ground Truth: The SPIDER data set provided the correct SQL queries corresponding to each natural language question, which served as the ground truth for comparison.
    
    \item Evaluation Metrics: The similarity between the generated SQL queries and the ground truth SQL queries was measured using the following metrics:
        \begin{itemize}
            \item Normalized Edit Distance: Quantifies the structural similarity between two SQL queries by computing the minimum number of edits required to transform one query into another.

            \item Embedding Matching: Captures semantic similarity by representing queries as vector embeddings and computing their cosine similarity.

            \item Fuzzy Matching: Evaluates textual similarity using string-matching techniques that tolerate minor variations.

            \item SQL Component Matching: This metric evaluates the individual components of SQL queries, such as table selection, column selection, and operation selection. Comparing these elements between the generated and correct SQL queries identifies which specific aspects of query generation are most prone to hallucination.

            \item Database Execution Comparison: This approach involves executing the generated and correct SQL queries on a PostgreSQL database and comparing their outputs. It provides a direct measure of query correctness by assessing the results' equivalence.
        \end{itemize}
        
\end{enumerate}
We calculated the percentage of query pairs for the first three similarity metrics where the similarity score exceeded a predefined threshold. This threshold allowed us to classify the generated SQL queries as similar or dissimilar to the ground truth. The similarity scores presented below in Figure 5 represent the percentage of queries within each document set's RAG configuration that were deemed similar based on the respective evaluation metric.

\begin{figure}
    \centering
    \begin{minipage}[t]{0.48\textwidth}
        \centering
        \includegraphics[width=\textwidth]{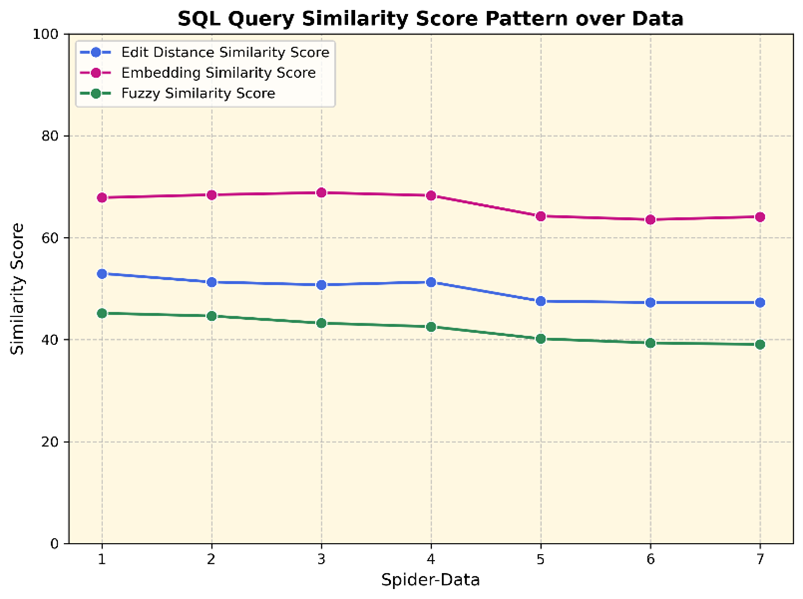}
        \caption{Percentage of queries marked as similar for each similarity metric across the different document sets.}
        \label{fig:sql_query_similarity_score_pattern}
    \end{minipage}%
    \hfill
    \begin{minipage}[t]{0.48\textwidth}
        \centering
        \includegraphics[width=\textwidth]{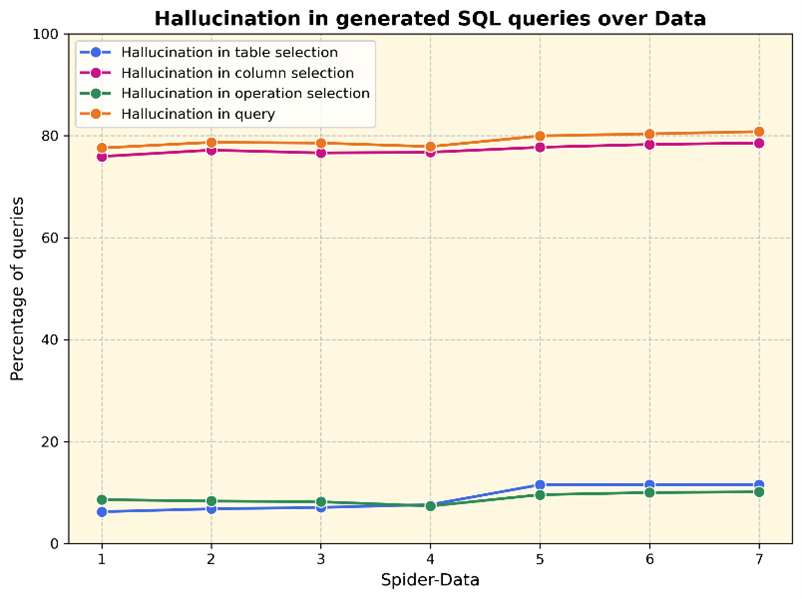}
        \caption{Percentage of Queries Exhibiting Mismatches Across Various SQL Query Components for Different Document Sets}
        \label{fig:hallucination}
    \end{minipage}
\end{figure}

Figure 6 illustrates the performance of the RAG + Text2SQL system across different document sets by analyzing mismatches in various components of the SQL queries. Specifically, it depicts the percentage of queries that experienced hallucination or mismatch between the generated SQL query and the correct SQL query for each component. This evaluation provides insights into which aspects of query generation were most prone to errors.

Next, in Figure 7, we plotted the percentage of pairs of SQL queries among the correct SQL queries provided by the SPIDER data set and the SQL generated by the Text2SQL model that, when executed, gave the same result in the database system. This will give us a measure of how the performance of the RAG+Text2SQL system changed across different Document sets.

\begin{figure}
    \centering
    \includegraphics[width=0.6\textwidth]{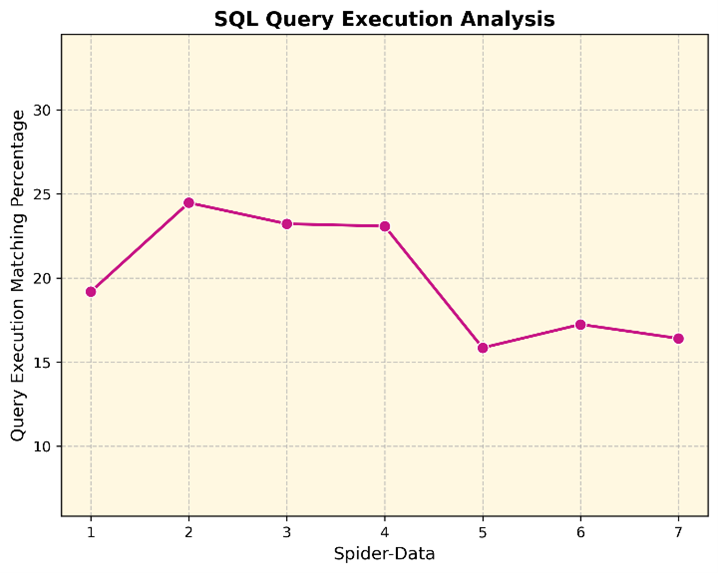}
    \caption{Percentage of queries giving same results on database execution across different document sets}
    \label{fig:sql_query_execution_analysis}
\end{figure}

The following observations, derived from the above plots and analysis, provide insights into the overall performance of the RAG+Text2SQL system across various document sets:
\begin{enumerate}
    \item Similarity Scores: Figure 5 illustrates the trends in similarity scores, revealing that the performance of the RAG + Text2SQL system declined as the content within the documents for RAG increased. This decline is attributed to the larger prompt size, which led to heightened hallucinations and greater dissimilarity between the correct SQL queries and the generated ones. While the similarity scores showed improvement from data set 1 to 4, the addition of textual descriptions in document set 5 caused a noticeable drop in all similarity metrics. This indicates that the increased content, although relevant, introduced complexities that impacted the model's accuracy.
    
    \item SQL Component Matching:  Figure 6 illustrates the percentage of queries exhibiting hallucinations in various components of SQL queries, highlighting an increase in hallucinations from document set 4 to document set 5. Among the components, hallucinations in table selection and operation selection were notably low, demonstrating that the Text2SQL model performed well in these areas.
    However, column selection showed significantly higher hallucinations. This was primarily due to the generated SQL queries including unnecessary ID columns from the tables, even when these columns were not required for the query. While these additional IDs did not affect the correctness of the query results, their inclusion was redundant despite explicitly instructing the model to avoid unnecessary columns; such behavior persisted, likely reflecting inherent limitations in the internal workings of the LLM.
    Addressing these issues may require larger or more advanced models than the 8B-parameter Text2SQL model used in this study. However, benchmarking alternative Text2SQL models is beyond the scope of this work. Hallucination rates increased across different document sets as the complexity grew.
    \item Database Execution Comparison: Figure 7 presents the percentage of queries that returned the same results upon database execution across different document sets. We compared the fetched results by executing both the correct SQL queries and the generated SQL queries. Our analysis revealed a significant decline in the quality of generated SQL queries between document sets 4 and 5. This drop in performance can be attributed to an increase in hallucinations, which was a direct consequence of the larger prompt sizes in the later document sets.
    \item Despite the overall improvement in RAG performance, hallucinations were observed in some cases as the document content grew excessively, highlighting the trade-off between context richness and prompt complexity.
\end{enumerate}

This evaluation highlights the nuanced interplay between document design and system performance in RAG + Text2SQL pipelines, emphasizing the importance of tailored document structures to achieve optimal results.

\section{Discussion}
This study explores the interplay between document size, quality, and the performance of Retrieval-Augmented Generation (RAG) + Text2SQL systems. The following points summarize our findings and highlight the critical factors influencing system performance:
\begin{enumerate}
    \item Impact of Document Content Size: Our experiments indicate that increasing the content size of documents benefits the RAG component independently but does not necessarily improve the overall performance of the RAG + Text2SQL system. Excessive content in documents may lead to diminishing returns or even negative effects on system performance.
    \item Performance Limitations with Increased Content: Contrary to intuition, expanding document content does not guarantee enhanced performance for the combined RAG + Text2SQL pipeline. This underscores the need to balance the document size to optimize the system as a whole.
    \item Effectiveness of One-Shot Examples: The utility of one-shot examples diminishes when documents already contain substantial content. In contrast, one-shot examples are more impactful when the document has limited textual information. This finding suggests that the role of examples should be strategically adjusted based on document quality and content density.
    \item Handling Real-World Document Challenges: Real-world documents used in RAG systems often contain noisy and redundant information. Cleaning and transforming such documents are essential for achieving better system performance. Testing the system with multiple document sets (2-3 variations) can provide a more robust understanding of its behavior and reliability.
    \item Insights from Figures 6 and 7: As illustrated in Figures 6 and 7, Document Set 4 exhibited the best overall performance. This set had minimal textual information and included examples related to data insertion, enabling the Text2SQL model to understand table structures better. These findings highlight the importance of optimizing document content to balance clarity and informativeness.
    \item Trade-Off Between RAG and Text2SQL Components:  While adding more textual information may enhance the RAG system’s ability to differentiate between tables, it can negatively impact the Text2SQL model. Larger prompts resulting from excessive document content tend to produce more hallucinated results, reducing the accuracy and reliability of the system.
\end{enumerate}

\section{Proposed Approach for Optimization}
To optimize the performance of RAG + Text2SQL systems, it is crucial to design document content thoughtfully. The following factors should be considered when preparing the documents:
\begin{enumerate}
    \item Number of Documents: When the document set is small or fewer SQL tables are involved, it is unnecessary to focus heavily on RAG. In such cases, documents without extensive textual content may suffice. Larger document sets with numerous tables, including detailed table descriptions and additional content, can enhance the RAG system’s ability to differentiate between tables, thereby improving its performance and reducing the selection of irrelevant tables.
    \item Quality of Tables/Documents: If the tables in the data set are highly similar (e.g., most tables belong to the same domain), a more powerful RAG system is required. Adding relevant content to the documents in such cases can help improve performance. If the document set comprises diverse and distinct tables, including a basic schema in the documents is often sufficient for RAG to retrieve relevant results.
    \item Quality of the Text2SQL Model: A high-performing Text2SQL language model (LLM) can handle more detailed content in the RAG system with lower risks of hallucinations. Therefore, when using a robust Text2SQL LLM, it is advantageous to include useful and descriptive content in the documents.
    \item Value of K in RAG: If the retrieval parameter in RAG is increased, the content of the documents should be adjusted accordingly. A higher value increases the size of the prompt passed to the Text2SQL model, which can influence the system’s efficiency and accuracy. In this case, smaller document sizes will be more suitable for LLMs to avoid longer prompt sizes.
    \item Dynamic Content Selection: Implementing a dynamic filtering mechanism can significantly reduce noise in the input. Allowing RAG to select only the most relevant parts of the schema document to pass to the Text2SQL model ensures the model focuses on essential information, improving overall system performance.
\end{enumerate}
In conclusion, optimizing document content in RAG-driven Text2SQL systems requires careful consideration of document size, quality, and alignment with the capabilities of the underlying models. These proposed strategies maximize system efficiency while minimizing errors, providing actionable insights for research and practical applications.

\section{Limitation and Future Work}
This section discusses the limitations of our experimental setup in evaluating the performance of the RAG + Text2SQL system across various document configurations and highlights potential directions for future research. The experiments in this study were conducted exclusively on the SPIDER (\cite{yu:2019}) data set, which, while being a widely recognized benchmark for Text2SQL systems, may not provide results that generalize to data sets with different structures, levels of complexity, or domain-specific characteristics. The schema documents and query patterns within SPIDER do not fully capture the diversity and variability that arise in real-world applications, limiting the broader applicability of our findings.

Another key limitation lies in the prompt length constraints of the Text2SQL model used in this study. For large or complex schemas, input truncation often occurs, leading to loss of context and consequently affecting system performance. Furthermore, the system assumes the existence of a predefined schema or structured database, which limits its adaptability when dealing with unstructured or semi-structured data. This reliance restricts the system’s application in scenarios where schemas are incomplete, evolving, or absent.  

Despite these challenges, our work opens several promising directions for future research. Investigating the scalability of the RAG + Text2SQL approach for large databases with extensive schemas is crucial, emphasizing improving computational efficiency, response times, and memory management to ensure practical viability in production settings. Fine-tuning the Text2SQL model on RAG-augmented inputs may enhance its ability to handle variations introduced by the retrieval system, thereby improving performance across diverse contexts.  

Efforts to mitigate hallucinations represent another critical avenue for exploration. Integrating domain-specific constraints, implementing confidence scoring mechanisms, or applying post-processing correction methods could prove beneficial. Additionally, advancing methods for dynamically generating schema documents tailored to specific query contexts hold significant promise. This approach would address challenges related to prompt length limitations while optimizing the balance between schema detail and simplicity, improving the system’s overall adaptability. Addressing these limitations and pursuing these research directions could considerably enhance the robustness, scalability, and practical applicability of RAG + Text2SQL systems in real-world settings.

\section{Conclusion}
This study has investigated the optimization of document design in Retrieval-Augmented Generation (RAG) systems for Text-to-SQL tasks. Key findings from the experiments have highlighted several critical insights into how document size, quality, and query generation accuracy can be balanced effectively. We found that providing structured and descriptive documents, incorporating syntax-corrected queries, and utilizing learning techniques such as zero-shot, one-shot, and two-shot learning are essential for improving the accuracy of SQL queries generated by models like Llama 3 SQLCoder. While incorporating table and column descriptions improved the performance of table and column selection, it also led to a higher frequency of syntax errors and hallucinations in certain cases. Moreover, we observed that balancing document size and content quality is vital for minimizing discrepancies in query results, with smaller deviations from the base queries achieved when using corrected syntax and simpler query structures.

The significance of these findings lies in their potential to guide the design of more effective RAG systems that can generate accurate SQL queries with minimal errors. Combining strategies such as dynamic content filtering, prompt construction, and adaptive document design based on query context can improve query generation accuracy and efficiency. These findings also emphasize the importance of reducing document size without sacrificing relevant details, thereby improving the model’s ability to generate accurate queries while minimizing unnecessary computational load. In conclusion, achieving an optimal balance in document design for RAG + Text-to-SQL systems involves a delicate trade-off between maintaining document quality and ensuring query accuracy by refining content filtering approaches and structuring relevant documents.

\bibliographystyle{unsrtnat}
\bibliography{balancing_content_size_in_rag_text2sql_system}

\end{document}